\documentclass[epj]{svjour}

\usepackage[super,compress]{cite}
\usepackage{graphicx}
\usepackage{amsmath,amssymb}

\begin{document}

\newcommand{\fr}{\frac}
\newcommand{\ct}{\cite}
\newcommand{\lb}{\label}
\newcommand{\ti}{\tilde}
\newcommand{\prm}{\prime}

\newcommand{\al}{\alpha}
\newcommand{\bt}{\beta}
\newcommand{\ka}{\kappa}
\newcommand{\la}{\lambda}
\newcommand{\La}{\Lambda}
\newcommand{\si}{\sigma}
\newcommand{\Si}{\Sigma}
\newcommand{\te}{\theta}
\newcommand{\Te}{\Theta}
\newcommand{\ga}{\gamma}
\newcommand{\ep}{\epsilon}
\newcommand{\ups}{\upsilon}
\newcommand{\Om}{\Omega}
\newcommand{\om}{\omega}
\newcommand{\ld}{\delta}
\newcommand{\CD}{\Delta}
\newcommand{\ot}{\otimes}
\newcommand{\vae}{\varepsilon}
\newcommand{\vap}{\varphi}
\newcommand{\kd}{\delta}

\newcommand{\app}{\approx}
\newcommand{\ra}{\rightarrow}
\newcommand{\eqv}{\equiv}
\newcommand{\di}{\diamond}
\newcommand{\tbf}{\textbf}
\newcommand{\mbf}{\mathbf}
\newcommand{\oln}{\overline}
\newcommand{\dgr}{\dagger}
\newcommand{\del}{\partial}

\newcommand{\sqr}{\square}
\newcommand{\nno}{\nonumber}
\newcommand{\noin}{\noindent}
\newcommand{\bit}{\bibitem}
\newcommand{\be}{\begin{equation}}
\newcommand{\ee}{\end{equation}}
\newcommand{\beqa}{\begin{eqnarray}}
\newcommand{\eeqa}{\end{eqnarray}}

\def\ds{\displaystyle}

%%%%%%%%%%%%%%%%%%%%%%%%%%%%%%%%%%%%%%%%%%%%%%%%%%%%%%%%%%%%%%%%%%%%

\title{Flat Galactic Rotation Curves from Geometry in Weyl Gravity}

\author{Cemsinan Deliduman\inst{1,2} \and O\u{g}uzhan Ka\c{s}\i k\c{c}\i \inst{1} \and Bar\i \c{s} Yap\i\c{s}kan \inst{1} }

\institute{Department of Physics, Mimar Sinan Fine Arts University, Bomonti 34380, \.{I}stanbul, Turkey, \\
\email{cemsinan@msgsu.edu.tr, oguzhan.kasikci@msgsu.edu.tr, baris.yapiskan@msgsu.edu.tr} \and
National Astronomical Observatory of Japan, 2-21-1 Osawa,
Mitaka, Tokyo 181-8588, Japan}

\abstract{
We searched for a resolution of the flat galactic rotation curve problem from geometry instead of assuming the existence of dark matter. We observed that the scale independence of the rotational velocity in the outer region of galaxies could point out to a possible existence of local scale symmetry and therefore the gravitational phenomena inside such regions should be described by the unique local scale symmetric theory, namely Weyl's theory of gravity. We solved field equations of Weyl gravity and determined the special geometry in the outer region of galaxies. In order to understand the effective description of gravitational phenomena, we compared individual terms of so called Einstein--Weyl theory and concluded that while the outer region of galaxies are described by the Weyl term, the inner region of galaxies are described by the Einstein--Hilbert term.
\PACS{{04.20.Cv}{}\and {04.20.Jb}{}\and {04.50.Kd}{}}
\keywords{Galactic rotation curves, alternative gravity.}
}
\maketitle

%%%%%%%%%%%%%%%%%%%%%%%%%%%%%%%%%%%%%%%%%%%%

\section{Introduction}

According to current cosmological paradigm, there is a ``dark side of the universe,''' which consists of Dark Energy and Dark Matter. Former is the proposed explanation to the observation of cosmic speed--up \cite{Perlmutter,Riess1,Riess2,Kowalski} and the latter is the proposed explanation to observation of flat rotation curves of galaxies \cite{JB,AB,KGB,PSS,BS}, gravitational lensing \cite{SKW,Dodelson}, and many other astrophysical and cosmological observations \cite{Bertone,Lyth}. Other than the existence of these dark components of the universe, contemporary $\Lambda$CDM cosmology ($\Lambda$ for Dark Energy and CDM for Cold Dark Matter) also assumes that Einstein's theory of gravity, i.e. general relativity, is valid at all length scales. Although $\Lambda$CDM cosmology successfully explains almost all the cosmological observations it has several important theoretical drawbacks. Even though diverse phenomenology is proposed to explain the nature of dark energy (see e.g. reviews \cite{PeRa,Sah,CST,FrTu,LLWW,ATsu}), to date there is no satisfactory model everyone agrees on. Ambitious experiments designed to observe elusive dark matter particles \cite{Tr,Berg,BFFP} gave negative results so far \cite{BHKK,ZA,LUX}. 

These theoretical problems led many physicists to seek alternative explanations to cosmological observations by challenging the assumption of $\Lambda$CDM cosmology on the theory of gravity: that the general relativity may not be valid in all length scales. Successful tests of predictions of general relativity in the Solar system (see e.g. \cite{SCar}) and the results of table--top experiments \cite{HNSS,LCP,DJK} on the range of validity of the Newtonian potential, which is the non--relativistic, weak field limit of general relativity, confirms that the general relativity is valid at least from a few microns to the size of the Solar system. However, we have no compelling evidence for the validity of the general relativity at scales larger than the scale of the Solar system. Gravity could be behaving fundamentally different in large scales, such as galactic or cosmic scales. This idea motivated the study of so called modified or alternative theories of gravity in recent years.

There are many diverse ways to modify Einstein's general relativity. Simplest of which is to have an arbitrary function of the scalar curvature as the Lagrangian density. Such theories are simply called the $f(R)$ theories of gravity \cite{Odintsov-rev,Sotiriou-rev,deFelice-rev,BC,DW}. The geometric f(R) term is motivated originally from renormalization of stress tensor operator of interacting quantum fields in curved space times (see, e.g., \cite{ParkerToms,BirrellDavies}). There are several works attacking the flat galactic rotational velocity curve problem in $f(R)$ gravity \cite{Sob06,Capoz,BHL}. In \cite{Sob06} and \cite{Capoz} the gravitational potential in the weak field limit is modified and this modification has either logarithmic \cite{Sob06} or power law \cite{Capoz} behavior. Whereas \cite{BHL} accepts the gravitational potential has the Newtonian form, but the ``mass'' required for the flat rotation curve phenomenology is declared to be ``geometric mass'' coming from a special solution for the metric. There are many other alternative gravity theories that propose gravitational or dynamical explanation to the observed Dark Matter phenomenology. Among them, one interesting subset of theories is the scalar-vector-tensor gravity theories: relativistic extension of Modified Newtonian Dynamics (MOND), so called Tensor-Vector-Scalar (TeVeS) theory of Bekenstein \cite{Bekenstein2004}; scalar-vector-tensor gravity (STVG) theory of Moffat \cite{Moffat2006}; Jordan-Brans-Dicke scalar-vector-tensor (JBD- SVT) theory of Ghaffarnejad \cite{Gh2008} are the notable examples. These theories predict a good correspondence with the experimental data \cite{Skordis2009,Br2005,Gh2019}.

In this article we will work in the Weyl gravity \cite{Weyl1,Weyl2}. The Weyl gravity has as its fundamental symmetry the local scale transformation, so called Weyl transformation, of the metric tensor. So it is not a modification of the general relativity, but it is truly an alternative theory written a few years after Einstein and Hilbert formulated general relativity. Weyl gravity do not have the correct Newtonian limit \cite{Sch}, and this prevented for almost a century for this theory to be taken seriously to describe the gravitational phenomena. Mannheim and Kazanas in the 1990's \cite{MaKa,Mann} considered the phenomenology of Schwarzschild--like solution to attack both of the dark problems. However their approach is claimed to be flawed \cite{Yoon} due to the fact that the Newtonian part of the gravitational potential is proportional to the second moment of the mass density, instead of the zeroth moment, which is the total mass of the gravitating object. We believe Yoon's critique is extremely important and thus the Weyl gravity approach to flat galactic rotation curve problem is far from complete. Taking into account Yoon's critique one needs to find new solutions of Weyl gravity which are exempt from such problems. This is the reason that we embarked on the present project and this aim determines also the scope of this paper.

Given the volume matter density of central bulge of a regular spiral galaxy as $\rho (r) = \rho_{ob}\, e^{-r/r_{ob}}$ with $r_{ob}$ being a characteristic length specific to the galaxy, the rotational velocities of stars located from a distance of nearly $2.2\, r_{ob}$ up to the edge of the galaxy have almost the same value \cite{Mann}. We will call this region of flat velocity curve \textit{the outer region} and the region of raising velocity curve (from $r=0$ to $r=2.2\, r_{ob}$) \textit{the inner region} following \cite{Mann}.  Newtonian gravity produced by the luminous matter can be seen solely responsible of the rising velocity curve from the center up to the peak value at $r=2.2\, r_{ob}$. After the peak point, however, the effect of Newtonian potential diminishes and related to that it is expected that the rotational velocities of stars would diminish depending on how much a particular star is away from the centre of the galaxy. Hence almost constancy of the rotational velocities beyond the peak value at $r=2.2\, r_{ob}$ points to the fact that something is different in the outer region of the galaxy. The current paradigm asserts the existence of non--luminous dark matter in that region. In this paper we would like to follow the idea that in fact gravity is fundamentally different in the outer region of the galaxy and perhaps beyond.

The scale independence of the rotational velocity in the outer region of galaxies could point out to a possible existence of local scale symmetry in such regions. Therefore the gravitational phenomena in the outer region of galaxies could be described by the unique local scale symmetric metric theory, namely Weyl's theory of gravity. Then we have to determine the special geometry in which all the objects rotating around a distant galactic center have the same rotational velocities. Here we solve the field equations of Weyl gravity and by finding non-trivial exact solutions, we determine special geometry of the outer region of galaxies in several cases of the geometry of the solid angle. At this point we would like to emphasize that we are not seeking a solution in the Newtonian limit in terms of a potential. We are claiming that the scale invariance of rotation curves in the outer region of galaxies could be due to the special form of the metric in that region, and this metric is an exact solution of the Weyl gravity. Thus we propose that the fundamental theory of gravity beyond the inner region of galaxies could be the Weyl gravity. In the last section, we are going to attempt to reconcile this proposal with the fact that the general relativity is very successful inside the Solar system and also in the inner galactic region.

We remark that the scope of the present paper is set at the level of \cite{Sob06,BHL} and \cite{MaKa}. This means that what we are looking in this paper is (are) some solution(s) to Weyl gravity that describes the outer regions of galaxies where the almost constant rotational velocities of the stars are observed irrespective to their distance from the galactic center. We do not attempt to do any fitting to actual data or in that respect we are not concerned with the problem of ``almost'' flatness of the rotation curves. These problems will be commented upon in the last section of this paper and planned to be thoroughly investigated as part of future research. We also remark that cosmological Dark Matter phenomenology is much harder to explain with an alternative gravity methodology. Thus, this article should be considered a part of an ongoing research project on astrophysical and cosmological implications of the Einstein-Weyl gravity.

The outline of this paper is as follows. In the next section we are going to review the basic structure of Weyl gravity and its field equations. Then in section (\ref{flat}) the tangential velocity of matter particles on circular orbits will be derived in terms of one of the metric functions. Afterwards in section (\ref{solutions}) four distinct solutions of Weyl gravity in the region of flat galactic rotation curves will be presented. In section (\ref{stability}) we will prove the stability of circular orbits in the backgrounds found in the previous section.  Finally, section (\ref{discussion}) will contain our discussion of the results and comments on the effective theory of the gravitational interactions.

%%%%%%%%%%%%%%%%%%%%%%%%%%%

\section{Weyl gravity and the field equations \label{theory}}

Action of Weyl gravity is given in terms of contraction of two Weyl tensors as
\be \label{Weyl_S}
S =\alpha \int d^{4}x \sqrt{-g} C_{\mu\nu\rho\sigma} C^{\mu\nu\rho\sigma}\, ,
\ee
where $C_{\mu\nu\rho\sigma}$ is the Weyl tensor, which is defined as the anti--symmetric part of the Riemann tensor, and $\alpha $ is a dimensionless parameter. As described in the introduction, this action is invariant under Weyl transformation, which is the local scale transformation of the metric tensor, given by
\be \label{Weyl_T}
g_{\mu\nu} \rightarrow \tilde{g}_{\mu\nu} = e^{\Omega (x)} g_{\mu\nu}\, .
\ee
The invariance of (\ref{Weyl_S}) under this transformation is due to the fact that Weyl tensor is invariant under local scale transformation:
\be
C_{\mu\nu\rho}^{\quad\ \sigma} \rightarrow \tilde{C}_{\mu\nu\rho}^{\quad\ \sigma} = C_{\mu\nu\rho}^{\quad\ \sigma}\, .
\ee
It can be shown \cite{Mann} that (\ref{Weyl_S}) is the unique metric action invariant under the Weyl transformation (\ref{Weyl_T}).

Since the Weyl tensor is the anti-symmetric part of the Riemann tensor, its self--contraction can be written in a special combination of self contractions 
of Riemann and Ricci tensors, and the scalar curvature. This expression is given by
\be
C_{\mu\nu\rho\sigma} C^{\mu\nu\rho\sigma} = R_{\mu\nu\rho\sigma} R^{\mu\nu\rho\sigma} 
- 2R_{\mu\nu} R^{\mu\nu} + \fr13 R^2\, .
\ee
In four dimensions there is another special combination of these terms, so called Gauss--Bonnet term,
\be
\mathcal{G} = R_{\mu\nu\rho\sigma} R^{\mu\nu\rho\sigma} - 4R_{\mu\nu} R^{\mu\nu} + R^2\, ,
\ee
which is topological invariant. This means that it does not contribute to the field equations and therefore only the part of 
$C_{\mu\nu\rho\sigma} C^{\mu\nu\rho\sigma}$ modulo Gauss--Bonnet term,
\be
C_{\mu\nu\rho\sigma} C^{\mu\nu\rho\sigma} 
= 2(R_{\mu\nu} R^{\mu\nu} - \fr13 R^2 )\quad \mathrm{(mod\, \mathcal{G})}
\ee
contributes to the field equations.

Thus the action of Weyl gravity that we are going to use is
 \be \label{Weyl_S2}
 S =2\alpha \int d^{4}x \sqrt{-g} (-\frac{1}{3}R^2+R_{\mu\nu} R^{\mu\nu}) \, .
 \ee
We are going to analyze the vacuum solutions of field equations of this theory in the constant rotation velocity galactic region.

We present the vacuum field equations of this theory as a combination of two terms. Contribution of the first term in the action (\ref{Weyl_S2}), that is $R^2$, to the field equations is given by
\be \label{Hmn}
H_{\mu\nu} = 2R\left( R_{\mu\nu} - \fr14 Rg_{\mu\nu}\right) +2\left( g_{\mu\nu}\Box - \nabla_\mu \nabla_\nu \right) R\; ,
\ee
and contribution of the second term in the action (\ref{Weyl_S2}), that is $R_{\mu\nu} R^{\mu\nu}$, to the field equations is
\beqa 
K_{\mu\nu} &=& \Box \left(R_{\mu\nu} + \fr12 g_{\mu\nu}R \right) 
- \nabla_\lambda \nabla_\mu R^\lambda _{\; \nu} 
- \nabla_\lambda \nabla_\nu R^\lambda _{\; \mu} \nno \\
&& +2R_{\mu\lambda}R^\lambda _{\; \nu} - \fr12 g_{\mu\nu}R^{\alpha\beta}R_{\alpha\beta}\; . \label{Kmn} 
\eeqa
Then the field equations of the Weyl gravity (\ref{Weyl_S2}) are given by
\be \label{Bach}
B_{\mu\nu}= -\frac{1}{3}H_{\mu\nu} + K_{\mu\nu}=0 \, ,
\ee
where $B_{\mu\nu}$ is called the Bach tensor \cite{Bach}. Since it is conformally invariant, trace of the Bach tensor vanishes.

%%%%%%%%%%%%%%%%%%%%%%%%%%%

\section{Flat rotation curve galactic region \label{flat}}

Since our aim is to determine the form of the metric in the constant rotation velocity galactic region we consider a spherically symmetric form for 
the metric:
\be
ds^2= - A(r)dt^2 +\frac{dr^2}{B(r)} +r^2 d\Omega_{k}^2 \, ,
\ee
where $ d\Omega_{k}^2 \equiv \frac{1}{1-kx^2}dx^2+(1-kx^2)dy^2 $ corresponds to two dimensional hyperbola, torus and sphere geometries of the solid angle, with $k$ taking values $-1,\, 0,\, 1$, respectively. In the sphere geometry we may use angular coordinates $\theta$ and $\phi$ to write 
$ d\Omega_{1}^2 = d\theta^2 + \sin^2 \theta\, d\phi^2 $. Since a spiral galaxy has a disc geometry, stars move on geodesics in the equatorial plane, and therefore in the sphere geometry we might take $\theta = \pi / 2$.

The kinematical condition that the tangential velocity of a circular orbit is constant gives us a constraint to determine one of these metric functions 
completely. Since we are analyzing timelike geodesics that stars follow we have $ds^2 < 0$, which allows us to write
\be \label{eom1}
1 = A\dot{t}^2 - \frac{\dot{r}^2}{B(r)} - r^2 \dot{\phi}^2\, , 
\ee
where dotted quantities represent derivative with respect to affine parameter $s$. For the case of circular timelike geodesics we further have $\dot{r}=0$, we then have
\be \label{eom2}
1 = A\dot{t}^2 - r^2 \dot{\phi}^2\, .
\ee
Derivative of this expression with respect to $r$ gives us further that
\be \label{eom3}
\frac12 A^{\prime}\dot{t}^2 = r \dot{\phi}^2\, .
\ee

Tangential velocity can be obtained from the expression given in \cite{Lan2} as
\be
v_c = \fr{dl}{ds} = \fr{rd\phi}{\sqrt{-g_{00}}dt}\, ,
\ee
where $l$ is the length measured on the geodesic. Using (\ref{eom3}), we obtain the tangential velocity as
\be \label{v2}
v_c^2 = \fr r2 \fr{A^{\prime}}{A}\, ,
\ee
where $A^{\prime} = \frac{dA}{dr}$. This relation was also obtained previously in \cite{Sob06,BHL}. 

Since we are interested in constant $v_c$ and since (\ref{v2}) is a first order differential equation in $A(r)$, it can be easily integrated. 
Setting $v_c^2 = w = \mathrm{constant}$ we obtain the form of the metric as
\be \label{met}
ds^2 = -\left( \frac{r}{r_c}\right)^{2w}dt^2+\frac{dr^2}{B(r)}+r^2 d\Omega_{2,k}\, ,
\ee
where $w = v_c^2\approx10^{-6}$ if we work with natural units in which $c=1$, and $r_c$ is an integration constant.

%%%%%%%%%%%%%%%%%%%%%%%%%%%

\section{Solutions in Weyl gravity \label{solutions}}

We have only one metric component left to determine. For that we need to solve only one field equation, and make sure that the other field equations 
are trivially satisfied by this solution. Due to the Bianchi identities and trace relations, all field equations can be written in terms of the 
``$rr$" component. So it's enough to solve just ``$rr$" component of the field equations\footnote{Please refer to appendix (\ref{Bmn}) for a short proof that, given the metric (\ref{met}), solutions of $B_{rr} =0$ are also the solutions of all the equations $B_{\mu\nu} = 0$.}. This equation can also be obtained by varying the action with respect to $B(r)$ after substituting the metric into the Lagrangian. Thus we determine the ``$rr$" field equation,
\be \label{equB}
BB''r^2+wBB'r-\frac{1}{4}B'^2r^2-B^2(w-1)^2+\frac{k^2}{(w-1)^2}=0\, ,
\ee
where $B'(r)$ is the derivative of $B(r)$ with respect to $r$. 

For the sake of the completeness we also write down the other components of the field equations. Due to the spherical symmetry ``$\theta\theta$'' and ``$\phi\phi$'' field equations are the same and given by
\beqa
B[B'''r^3+3wB''r^2+(3w-2)B'r]+\frac{1}{2}B''B'r^3 \nno \\
\qquad +\frac{w}{2}B'^2r^2-2(w-1)^3B^2+\frac{2k^2}{w-1}=0\, ,
\eeqa
whereas the ``$tt$'' field equation is
\beqa
B[B'''r^3+(2w+1)B''r^2-(w^2-4w+2)B'r] \nno \\
\qquad +\frac{1}{2}B''B'r^3 +\frac{3w-1}{4}B'^2r^2 \nno \\
\qquad -(w-1)^3B^2 +\frac{k^2}{w-1}=0\, .
\eeqa
These equations are satisfied by all the solutions of equation (\ref{equB}) presented in this paper.

The simplest solution of equation (\ref{equB}) is the obvious constant solution, $B(r)=C$, with
\be
C=\frac{k}{(w-1)^2}
\ee

This solution is exactly equivalent to the ``\textit{constant} $\lambda$" solution described in \cite{BHL}, except the fact that it is obtained as a 
solution of the Weyl gravity instead of the $f(R)$ gravity. In fact, this was the only solution that could be obtained exactly in \cite{BHL}. In the 
present paper we are going to show that there are exact solutions of Weyl gravity in which $B(r)$ is not 
a constant. These solutions are important in two respects: firstly they are newly found exact solutions to Weyl Gravity, which is important from mathematical physics point of view, and secondly $k=1$ sphere geometry solution is our main contribution to the idea that the galactic flat rotation curves can be explained via geometry instead of dark matter. Even though only solution with $k=1$ sphere geometry has physical meaning and can be used as the metric of outer region of galaxies, here we present all the solutions for the sake of completeness.

%%%%%%%%%%%%%%%%%%%%%%%%%%%
  
\subsection{Solution specific to torus $T^{2}$ geometry \label{solutions1}}

We first find an analytical solution for the  2-torus $T^{2}$ $(k=0)$ case. We perform a transformation $ B(r)=[b(r)]^n $ after which 
the equation (\ref{equB}) we want to solve becomes 
\begin{equation}
b[b''r^2+wb'r-\frac{1}{n}(1-w)^2b]+(\frac{3n}{4}-1)(b')^2r^2=0,
\end{equation} 
which has a simple form. After a choice $n=\frac{4}{3}$, we get a second order linear differential equation which can be solved easily,
\begin{equation}
b''r^2+wb'r-\frac{3}{4}(1-w)^2b=0\, .
\end{equation}   
To solve this equation we make a solution ansatz $b(r)=r^\alpha $. Then the characteristic equation 
\begin{equation} \label{character}
  \alpha^2 -(1-w)\alpha-\frac{3}{4}(1-w)^2=0 ,
\end{equation}
has to have real solutions since the discriminant is positive. The general solution for $b(r)$ is given by
\begin{equation}
  b(r)=C_1 r^{\alpha_{1}}+C_2 r^{\alpha_{2}}\, ,
\end{equation}
where $C_{1,2}$ are the integration constants, $ \alpha_{1}=\frac{3}{2}(1-w)$ and $\alpha_{2}=-\frac{1}{2}(1-w)$ are the 
roots of (\ref{character}). Thus from the ``$rr$" field equation we find the solution for $B(r)$ as 
\begin{equation}
  B(r)=(C_1 r^{\alpha_1}+C_2 r^{\alpha_2})^{\frac{4}{3}}\, .
\end{equation}
After inserting the expressions for $\alpha_1$ and $\alpha_2$ into this solution one gets
\begin{equation}
  B(r)= r^{2(1-w)} (C_1 +C_2 r^{-2(1-w)})^{\frac{4}{3}}\, ,
\end{equation}  
which is independent of $n$. Therefore the metric that satisfies the field equations is   
\begin{equation}
 ds^2 = -\left( \frac{r}{r_c}\right)^{2w}dt^2+\frac{r^{2(w-1)} dr^2}{(C_1 +C_2 r^{2(w-1)})^{\frac{4}{3}}}
 +r^2 (dx^2+dy^2)\, .
\end{equation}
  
%%%%%%%%%%%%%%%%%%%%%%%%%%%

\subsection{Solution valid for sphere $S^{2}$ and hyperbolic plane $H^{2}$ geometries\label{solutions2}}

In order to simplify the equation (\ref{equB}) we perform the following transformation:
  \begin{equation} \label{var1}
  B(r)=\frac{2k}{(1-w)^2}r^{2(1-w)}F(r),
  \end{equation}
which yields
  \beqa
  F(r)\left[r^{2}F''(r)+(3-2w)rF'(r)\right]-\frac{r^2}{4}F'(r)^{2} \nno \\
  \qquad +\frac{(1-w)^2}{4r^{4(1-w)}}=0.
  \eeqa
After a change of variable, $r=z^{1/(1-w)}$, we get
  \begin{equation}
  F(z)\left[z^{2}F''(z)+3zF'(z)\right]-\frac{z^2}{4}F'(z)^{2}+\frac{1}{4z^4}=0.
  \end{equation}
Now multiplying by $2z^4F'F^{-3/2}$, we obtain
  \beqa
  2z^{6}F'F''F^{-1/2}+6z^{5}F'^2F^{-1/2}-\frac{1}{2}z^{6}F'^3F^{-3/2} \nno \\
  \qquad +\frac{1}{2}F'F^{-3/2}=0.
  \eeqa
This equation can be written as a total derivative,
  \begin{equation}
  \frac{d}{dz}\left(z^{6}F'^2F^{-1/2}-F^{-1/2}\right)=0\, ,
  \end{equation}
from which we can easily solve for the first derivative of $F(z)$ with respect to $z$:
  \begin{equation}
  F'(z)=\left(1+\sqrt{16C_1/3}\sqrt{F(z)}\right)^{1/2}\frac{1}{z^3},
  \end{equation} 
where we have chosen the integration constant as $\sqrt{16C_1/3}$ for later convenience. To integrate this equation we make another useful definition,
  \begin{equation}\label{var2}
    v(z)=1+\sqrt{16C_1/3}\sqrt{F(z)}
  \end{equation}
with
  \be
    dF=\frac{3}{8C_1}(v-1)dv \, .
  \ee
Then the integration of
  \be
    \frac{dF}{\left(1+\sqrt{16C_1/3}\sqrt{F}\right)^{1/2}}=\frac{dz}{z^3} 
  \ee
becomes a cubic algebraic equation of $\sqrt{v}$ given by
  \begin{equation}\label{cebir} 
  \left( \sqrt{v} \right)^3-3\sqrt{v}=2q(z)\, ,
  \end{equation} 
where $q(z)=-\frac{C_1}{z^2}+C_2$. Using Cardano's method for this type of cubic equations we further write 
  \be
    \sqrt{v}=a+b,
  \ee
and then substituting this into (\ref{cebir}) yields
  \be
    a^3+b^3+3(ab-1)(a+b)-2q=0.
  \ee
If we now take $ab=1$, then 
  \be
    a^3+b^3-2q=0.
  \ee
This way we reach a second order equation 
  \be
    x^2-2qx+1=0,
  \ee
where $x$ stands for both $a^3$ and $b^3$. This equation is described by the relations,
  \be
    a^3+b^3=2q, \ a^3b^3=1.
  \ee
Then, the three solutions of (\ref{cebir}) are given by
  \beqa
    v_{1}(z) &=& (h+h^{-1})^2\, , \nno \\
    v_{2}(z) &=& (e^{i4\pi/3} h + e^{i2\pi/3} h^{-1})^2\, , \nno \\
    v_{3}(z) &=& (e^{i2\pi/3} h + e^{i4\pi/3} h^{-1})^2\, ,
   \eeqa
where  $h\eqv \Big[q(z)+\sqrt{q(z)^2-1}\Big]^{1/3}$.
  
As seen above, there are two kinds of solutions for $v(z)$. If $h$ is real, there are one real and two complex solutions, 
and if $h$ is complex, then all solutions are real.

Using the relations (\ref{var1}) and (\ref{var2}), and then transforming the coordinate back to $ r $ from $z$, all metric function solutions are found to be
  \begin{equation}  \label{B}  
   B_{i}(r)=\frac{3kr^{2(1-w)}}{8(1-w)^2C_1} (1+u_{i}(r))^2  \qquad (i=1,2,3)\, ,
  \end{equation}
with 
  \beqa
    u_{1}(r) &=& h(r)^2+h(r)^{-2}\, , \nno \\
    u_{2}(r) &=& \frac{1}{2}[(-1+i\sqrt{3})h(r)^{2} - (1+i\sqrt{3})h(r)^{-2}]\, , \nno \\
    u_{3}(r) &=& \frac{1}{2}[(-1+i\sqrt{3})h(r)^{-2} - (1+i\sqrt{3})h(r)^{2}]\, ,
   \eeqa
where $$ h(r)=\Big( -\frac{C_1}{r^{2(1-w)}}+C_2+ \sqrt{(-\frac{C_1}{r^{2(1-w)}}+C_2)^2-1} \Big)^{\frac{1}{3}}.$$
These solutions describe the spacetimes with two dimensional hyperbola and sphere geometries for $k=-1$ and $k=1$, respectively.
  
This way we determined, as $k=1$ solution of the Weyl gravity, the general geometry that allows stars moving in this geometry to have 
the same rotational velocity independent of their individual distance from the galactic center. According to the observations,
such a region starts from the edge of the central bulge at about $r\approx 2.2 r_{ob}$ to the edge of the galaxy \cite{Mann}.
Therefore our $k=1$ solution only describe this \textit{outer region} of the galaxy. 
The geometry of the central bulge, i.e. the \textit{inner region} of the galaxy,
is described by the Einstein's theory of gravity. How  the two theories could be reconciled at the border of inner and 
outer regions is commented upon in the final section.

Before finishing this subsection we would like to emphasize that from cosmological viewpoint Schwarzschild geometries should be embedded in the metric that describe cosmological dynamics (e.g., Friedmann-Robertson-Walker or de Sitter metric). This was done in the case of general relativity in \cite{Einstein1945}. Similarly, our solution (\ref{B}) should be considered to be valid in a limited region: in the outer region of galaxies up to the scale of a typical galaxy cluster ($\sim 10\ Mpc$). That this solution would remain static in an expanding universe is guaranteed by the existence of conformal gravity Birkhoff theorem \cite{Riegert1984}.

%%%%%%%%%%%%%%%%%%%%%%%%%%%

\section{Stability of circular orbits \label{stability}}

One non--trivial check of the proposed geometrical solution is the stability of the circular orbits in that geometry. Circular orbits must be stable in the outer region of galaxies to be consistent with the observations. We do the following analysis to prove the stability of the circular orbits.

For a  static and spherical symmetric spacetime metric, the geodesic equation of a massive particle in the planar motion can be found using
some constants of motion that come from Killing symmetries. For the metric,
\begin{equation}  
ds^{2}=-A(r)dt^{2}+\frac{1}{B(r)}dr^{2}+r^2 d\theta^2+r^2\sin^2\theta d\phi^2, 
\end{equation}
the metric functions, $A(r)$ and $B(r)$, depend only on the radial coordinate $r$ and not the other coordinates $(t,\, \theta,\, \phi )$ in the sphere geometry, therefore there are two Killing vectors, which are given by
\be
K = K^\mu\partial_\mu = \fr\partial{\partial t}\, \quad \mathrm{and} \quad L = L^\mu\partial_\mu = \fr\partial{\partial \phi} \, .
\ee
In components form these are
\be
K_\mu = (-A(r),\, 0,\, 0,\, 0)\, \quad \mathrm{and} \quad L_\mu = (0,\, 0,\, 0,\, r^2\sin^2\theta)\, .
\ee
Then the geodesic equations are equivalent to the statements that
\be
K_\mu \fr{dx^\mu}{ds} = constant \quad \mathrm{and} \quad 
L_\mu \fr{dx^\mu}{ds} = constant \, ,
\ee
which translate into
\be
A(r)\dot{t} = E\, \quad \mathrm{and} \quad r^2\sin^2\theta \dot{\phi} = L \label{EL} \, .
\ee
From the metric one obtains the equation for the radial coordinate in the equatorial plane ($\theta=\frac{\pi}{2}$) as 
\begin{equation} \label{1}
 \dot{r}^2=B(r)\left(\frac{E^2}{A(r)}-\frac{L^2}{r^2}+\epsilon\right) \, ,
\end{equation}
where $\epsilon=-1$ is for timelike, $\epsilon=1$ is for spacelike and $\epsilon=0$ is for null geodesics.
On the other hand, for radial components in the equatorial plane ($\theta=\frac{\pi}{2}$), geodesic equations give us
\begin{equation} \label{2}
\ddot{r}=\frac{B'}{2B}\dot{r}^2+B\left(\frac{L^2}{r^3}-\frac{A'E^2}{2A^2}\right) \, ,
\end{equation}
where we used the constants of motion $ (E,L)$ defined above (\ref{EL}). If we define effective potential $V_{eff}$ as 
\begin{equation}
V_{eff}\equiv B(r)\left(\frac{E^2}{A(r)}-\frac{L^2}{r^2}+\epsilon\right)\, ,
\end{equation}
then one can rewrite the equation (\ref{2}) as 
\begin{equation} \label{3}
\ddot{r}=\frac{V'_{eff}}{2}.
\end{equation}
For circular orbits we have $\dot{r}=0$ and $\ddot{r}=0$. Therefore we have two conditions on the effective potential as $V_{eff}=0$ and $V'_{eff}=0$ at the radius of the circular orbit, $r=R$. These give us the constants of motion  $ (E,L)$ in terms of the radius of circular orbit:
\begin{align}
E^2=\frac{-2\epsilon A^2}{2A-RA'}\, , \\
L^2=\frac{-\epsilon A' R^3}{2A-RA'}\, .
\end{align}
If we make a small perturbation, $r=R+\delta$, to the circular orbit, we obtain 
\begin{equation}
\ddot{\delta}=\frac{V''_{eff}}{2}\delta.
\end{equation}
Therefore the stability condition for circular orbits becomes  $V''_{eff}\leq0$.
Calculating $V''_{eff}$ at $r=R$ we get
\begin{equation}
V''_{eff}=\frac{-2B\epsilon}{A(2A-RA')}(2A'^2-AA''-3A'A/R).
\end{equation}
For $A=r^\omega/r_c^\omega$ ($r_c,\omega > 0$) we find $V''_{eff}=2B\epsilon \omega/R^2$. For a massive particle $V''_{ef}=-2B \omega/R^2$, where $B>0$. This proves the stability of the circular orbits in the geometry described by (\ref{met}). The same result  can also be found using Rayleigh's criterion \cite{Letelier} or Lyapunov exponents  \cite{Cardoso}. 

%%%%%%%%%%%%%%%%%%%%%%%%%%%

\section{Discussion of results and conclusions \label{discussion}}

In this article we searched a resolution of flat galactic rotation curve problem from geometry instead of assuming the existence of dark matter. The first observation we made is that the scale independence of the rotational velocity in the outer region of galaxies could point out to a possible existence of local scale symmetry inside such regions. Therefore the gravitational phenomena in the outer region of galaxies should be described by the unique local scale symmetric metric theory, namely Weyl's theory of gravity. Then we had to determine the special geometry in which all the objects rotating around a distant galactic center have the same rotational velocity. We solved field equations of Weyl gravity and determined the special geometry of the outer region of galaxies in several cases of the geometry of the solid angle. 

The fact that the phenomenology of raising rotation velocity curves in the central bulge of galaxies is explained very successfully by general relativity brings us to the natural question of how these two theories can be reconciled inside the galaxy or in that respect in any other system strongly dependent on gravitational physics. To resolve this issue in our model, we propose that the true theory of gravity is the combination of general relativity and the Weyl gravity with an action given by
\be \label{EW}
S = \fr{1}{2\kappa} \int d^{4}x \sqrt{-g}
\left[ R + \tilde{\alpha}\, C_{\mu\nu\rho\sigma} C^{\mu\nu\rho\sigma}\right]\, ,
\ee
where $\kappa = 8\pi G / c^4$ is Einstein's constant. Since the self--contraction of Weyl tensor has already mass dimension $4$, the coefficient $\tilde{\alpha}$ here has dimension of $[length]^2$ and related to $\alpha$ of (\ref{Weyl_S}) by $\tilde{\alpha} = 2\kappa\alpha$.

The relative values of the Einstein--Hilbert term $R$ and the Weyl term $\tilde{\alpha}\, C_{\mu\nu\rho\sigma} C^{\mu\nu\rho\sigma}$ would 
determine whether one is dominant in a specific setting or both of them should be considered on an equal footing \cite{Psaltis2008, Phillips}. We assume a galaxy to be made of pressureless matter (stars) with a mass density profile is given by exponential spheroid model in the bulge and the exponential disk model in the disk\footnote{There are several other models which are also often used in practice, such as e.g. de Vaucouleurs and S\'ersic laws for the bulge; Kuzmin, Mestel and Miyamoto-Nagai models for the disk. More details about this topic could be found in \cite{Binney2008, Sofue2013} .}. The relative values of the Einstein--Hilbert and the Weyl terms could be calculated \cite{Eksi2014} in the framework of general relativity via their expressions in terms of hydrodynamical quantities given by
\beqa \label{R-C2}
\qquad \qquad R (r) &=& \frac{8\pi G}{c^2} \rho (r) \, , \\
C_{\mu\nu\rho\sigma} C^{\mu\nu\rho\sigma} (r) &=& \frac{48 G^2}{c^4} \left(\fr{m (r)}{r^3} - \frac{4\pi}{3} \rho (r)\right)^2 \, , \nno
\eeqa
where $\rho (r)$ is the volume density, and $m(r)$ is the mass contained up to radial distance $r$ from the center of the galaxy. 

Under the assumption of exponential spheroid bulge plus exponential disk model for a regular spiral galaxy it is not hard to determine the values of $R$ and $C_{\mu\nu\rho\sigma} C^{\mu\nu\rho\sigma}$ by using the above relations (\ref{R-C2}). As explained in \cite{Mann}, in the maximum disk prescription the rotation velocity data is well explained up to the border of the central bulge by the Newtonian theory. Therefore for our proposal to make sense, the Weyl term, $\tilde{\alpha}\, C_{\mu\nu\rho\sigma} C^{\mu\nu\rho\sigma}$, should dominate over the Einstein--Hilbert term, $R$ in the outer region of galaxy. This is shown to be true for a typical galaxy in appendix (\ref{A}). We interpret this to be a system dependent turning on/off of two parts of the Einstein--Weyl theory. Here we emphasize that the full solution of Einstein--Weyl theory is not needed. In different regimes different terms are dominant, thus in different regimes only solutions to dominant terms are relevant. When matter density is high one should use Einstein-Hilbert term and this theory decides its range of applicability through relations (\ref{R-C2}). At low matter densities Weyl term dominates and then all the geodesics are determined according to a solution of Weyl gravity. This is the essence of the approach followed here. In that respect it is similar to the  approach taken in \cite{Maeder2017b}.

It is pretty clear that the relative value of Einstein--Hilbert and Weyl terms depends also on the value of $\tilde{\alpha}$. This coefficient and the Weyl term it multiplies has different meanings depending on the context the theory is utilized. In the ultraviolet, the Weyl term is needed as a counterterm to make the theory of general relativity renormalizable \cite{Stelle1,Fradkin,Holdom}. Then $\tilde{\alpha}$ is a dimensionful coupling that depends on scale \cite{Holdom,Hooft} according to a $\beta-$function which turns out to be negative. Possible ultraviolet effects of the Weyl term is confined inside the central black hole of the galaxy and it is not relevant to the dynamics of the stars near the center of the galaxy. 

In the infrared, however, the same Weyl term could be considered as a high--derivative correction to general relativity, very similar in concept as the chiral Lagrangian of QCD \cite{Holdom}, but now the coefficient of Weyl term has different meaning than ultraviolet. In this respect the Einstein--Weyl gravity is an effective description of gravitational phenomena in the IR (see \cite{Holdom} and \cite{Donoghue2017} and references therein). According to \cite{Holdom} the couplings of higher derivative terms (one of them is the Weyl term) run due to perturbative corrections in this effective theory. This running of the coupling constants (one of them is $\alpha$) may not depend only on the energy, but there could also be dependence on ``a second expansion parameter - roughly the integrated curvature -.'' \cite{Donoghue2017} Astrophysical scales are obviously in the infrared and cosmological scales are in the far infrared. In such scales, we interpret $\tilde{\alpha}$ as a running coupling constant which could have different values depending on the characteristics of the system at that scale. Thus we could expect different values of $\tilde{\alpha}$ in the Solar system, in the inner region of galaxy, or in the outer region of galaxy. It is beyond the scope of this work to calculate the exact values of $\tilde{\alpha}$ in those regions. In this work we only aim to show that it is possible to have dominance of different terms of Einstein--Weyl gravity in accordance with the observed phenomenology. Please refer to appendix (\ref{A}) for the values of Einstein--Hilbert term, the Weyl term and the coupling constant $\tilde{\alpha}$ in various regions.

Similarity with the standard model of particles could be more than what is mentioned in the previous paragraph, though. In recent years, some interesting works published \cite{Shaposhnikov,Chankowski,Gorsky,Ghilencea} on the resolution to hierarchy problem by constructing asymptotic conformal symmetric, as an alternative to supersymmetric, theory of particles whose conformal symmetry would be broken by Higgs sector. In spirit we have a similar approach. We have conformal symmetric fundamental action in UV and in the far IR whose conformal symmetry is broken by Einstein-Hilbert term at the intermediate scales.

At this point we would like to point out similarities with some other works in the literature that have an extra characteristic length in addition to the Schwarzschild radius. General relativity, having equations of motion of second order in derivatives, have only one characteristic length, namely the Schwarzschild radius which is related to a property of the gravitational system, namely the mass of the object that creates the Schwarzschild geometry. In contrast, $f(R)$ gravity have equations of motion of fourth order in derivatives, and thus a further characteristic length emerges in $f(R)$ gravity due to existence of further conserved quantities obtained from the Noether symmetries \cite{Cap2007,Bernal2011}. This extra characteristic length also depends on the properties of the gravitational system \cite{Bernal2011,Jovanovic2016, Cap2017} and can be related to some phenomenological effective property of the system: in the case of spiral galaxies ``new gravitational radius'' can be related to MOND acceleration constant \cite{Bernal2011,Cap2017} and in the case elliptical galaxies it can be related to the galaxy effective radius which is one of the parameters of the fundamental plane of an elliptical galaxy \cite{Binney2008,Jovanovic2016}. Therefore, it is concluded that gravity could not be the same interaction at every scale. The enormous success of general relativity below Solar system scales is not an indicator of its validity at the other scales. Additionally, the same gravitational theory might show different phenomenological effective behavior at different scales. $f(R) = R^n$ theories, having two characteristic lengths, behave very differently at the scales corresponding to these lengths \cite{Bernal2011,Jovanovic2016, Cap2017}. This is also exactly the case here: Einstein-Weyl theory also have fourth-order field equations and thus it should have two characteristic lengths, which can be obtained by the Noether symmetry method of \cite{Cap2007}. Since Schwarzshild geometry is also a solution of Einstein-Weyl theory, one of the characteristic lengths is Schwarzschild radius. We expect that the additional characteristic length will be related to the constant $\alpha$ in front of the Weyl term in (\ref{EW}). Note that $\sqrt{\alpha\kappa c}$ has the dimension of length. Thus, running of $\alpha$ depending on the characteristics of the system, as described above, is effectively equivalent to gravity behaving differently at the new characteristic scale, which in turn depends on the properties of the gravitating system. We plan to investigate this connection quantitively in a future work. Using Noether symmetry method of \cite{Cap2007} we plan to derive the new gravitational radius in the Einstein-Weyl theory and check its relation to $\alpha$ and the phenomenological effective length scales of spiral and elliptical galaxies.

A further check of astrophysical relevance of our approach will be theoretical explanation of the gravitational lensing data of galaxy clusters and elliptical galaxies (see \cite{Dodelson} and references therein). In this respect, study of gravitational lensing in the Weyl gravity is very important and warrants a separate publication, which is currently under preparation. 

We also plan to investigate possible effect of Einstein--Hilbert term on the breaking of the scale symmetry in the outer regions of the galaxies. The rotation curves in the outer regions of galaxies are not exactly flat and this points out slight violation of the scale symmetry on which the Weyl theory of gravity is based. This is another indication that the resolution of flat galactic rotation curve problem cannot be found in pure Weyl gravity and supports our proposal that the full theory should be the Einstein--Weyl gravity. 

%%%%%%%%%%%%%%%%%%%%%%%%%%%

\section*{Acknowledgments}

C.D. thanks Toshitaka Kajino and the COSNAP group at NAOJ for discussions. We acknowledge support from the Turkish Council of Research and Technology (T\"{U}B\.{I}TAK)  with the project number 114F239 and the COST Action MP1405 (QSPACE). 

%%%%%%%%%%%%%%%%%%%%%%%%%%%

\appendix

\section{Values of $R$, $C_{\mu\nu\rho\sigma} C^{\mu\nu\rho\sigma}$ and $\tilde{\alpha}$} \label{A}

In this appendix we calculate values of $R$ and $\tilde{\alpha}C_{\mu\nu\rho\sigma} C^{\mu\nu\rho\sigma}$ both at the onset of the flat rotation curves region and at the edge of the galaxy to show that in the outer region of the galaxy the Weyl term dominates over the Einstein--Hilbert term. Since the coefficient of the Weyl term, $\tilde{\alpha}$, in (\ref{EW}) has dimension of $[length]^2$ we set $\tilde{\alpha} = r_c^2$ with $r_c$ being a scale length. The calculation in the outer region thus determines an order of magnitude value for 
$\tilde{\alpha}$ in a typical galaxy. With the same value of $\tilde{\alpha}$ we also show that in the inner region of the galaxy the Einstein--Hilbert term dominates over the Weyl term. The value of $\tilde{\alpha}$ in the inner region could be less than its value in the outer region of galaxy, in which case the dominance of  the Einstein--Hilbert term would be even more pronounced. Inside the Solar system the value of $\alpha$ was calculated some time ago in \cite{Mannheim2007} and it was shown that the Solar system tests of gravity are not affected by the existence of the Weyl term. 

We use the realistic mean values for a typical galaxy as given in \cite{Sofue2015}. For the bulge we assume exponential spheroid model with volume density given by $\rho (r) = \rho_{ob} \exp (-r/r_{ob})$. Mass of the bulge is given by $M_b = 8\pi r_{ob}^3 \rho_{ob}$. We take onset of outer region to be $r_b = 2.2 r_{ob}$ \cite{Mann} as explained in the introduction. Given the mean values for the bulge mass, $M_b = 2.3 \pm 0.4 \times 10^{10}\ M_\odot $, and the bulge scale radius, $r_{ob} = 1.5 \pm 0.2\, kpc$ of a typical galaxy (see table 2 of \cite{Sofue2015}), we calculate $R$ and $C_{\mu\nu\rho\sigma} C^{\mu\nu\rho\sigma} = C^2$ at $r = r_b$ from the relations (\ref{R-C2}) to be
\beqa
R (r_b) &=& \frac{G}{c^2} \frac{M_b}{r_{ob}^3}e^{-r_b/r_{ob}} \, , \\
C^2 (r_b) &=& \frac{48 G^2}{c^4} \frac{M_b^2}{r_{ob}^6}\left(\left(\frac{r_{ob}}{r_b}\right)^3 - \frac16 e^{-r_b/r_{ob}}\right)^2 \, . \nno
\eeqa
Thus the ratio of $r_c^2 C^2$ to $R$ at $r = 2.2 r_{ob}$ is about $80$ for $r_c$ on the order of $10\, Mpc$.

We now repeat this calculation at the edge of galaxy with the mean radius of a typical galaxy given by $r_g = 16 r_{od}$ \cite{Mann} with $r_{od}$ being the disk scale radius. For the disk we assume exponential disk model with volume density given by $\rho (r,z) = \rho_{od} \exp (-r/r_{od}-|z|/z_{od})$ in cylindrical coordinates. Vertical profile scale radius is typically $z_{od} \approx 0.1 r_{od}$. Mass of the disk is then given by $M_d = 4\pi z_{ob}r_{ob}^2 \rho_{ob}$. Given the mean values for the disk mass, $M_d = 5.7 \pm 1.1 \times 10^{10}\ M_\odot $, the total baryonic mass, $M_g = M_b + M_d = 7.9 \pm 1.2 \times 10^{10}\ M_\odot $, and the disk scale radius, $r_{od} = 3.3 \pm 0.3\, kpc$ of a typical galaxy (see table 2 of \cite{Sofue2015}), we calculate $R$ and $C_{\mu\nu\rho\sigma} C^{\mu\nu\rho\sigma} = C^2$ at $r = r_g$ from the relations (\ref{R-C2}) to be
\beqa
R (r_g) &=& 20\frac{G}{c^2} \frac{M_d}{r_{od}^3}e^{-r_g/r_{od}} \, ,  \\
C^2 (r_g) &=& \frac{48 G^2}{c^4} \frac{M_d^2}{r_{od}^6}\left(\left(\frac{r_{od}}{r_g}\right)^3\frac{M_g}{M_d} - \frac{10}3 e^{-r_g/r_{od}}\right)^2 \, . \nno
\eeqa
Thus the ratio of $r_c^2 C^2$ to $R$ at $r = r_g$ is about $20$ for $r_c$ on the order of $10\, Mpc$.

We note that value of $r_c$ is much larger than radius of any galaxy. In fact it is on the order of radius of typical galaxy cluster. This means that starting from the edge of central bulge of a galaxy up to the radius of a typical cluster, of which it is a member of, the Weyl term dominates the Einstein--Hilbert term and therefore the geometry should be described by the Weyl gravity. 

In the inner region of galaxy, however, we need to show just the opposite, that the Einstein--Hilbert term dominates the Weyl term and therefore the geometry should be described by general relativity. We take the value of $\tilde{\alpha}$ in the inner region of the galaxy to be at the same order of magnitude as its value in the outer region of the galaxy. 
We calculate $R$ and $C_{\mu\nu\rho\sigma} C^{\mu\nu\rho\sigma} = C^2$ at $r = 0.2 r_{ob}$ from the relations (\ref{R-C2}) to be
\beqa
R (r_b) &=& \frac{G}{c^2} \frac{M_b}{r_{ob}^3}e^{-r/r_{ob}} \, , \\
C^2 (r_b) &\approx& \frac{48 G^2}{c^4} \frac{M_b^2}{r_{ob}^6} e^{-r/r_{ob}} \left( \frac1{6!} \frac{r}{r_{ob}} \right)^2 \, . \nno
\eeqa
Thus the ratio of $R$ to $r_c^2 C^2$ at $r = 0.2 r_{ob}$ is about $10$ for $r_c$ on the order of $10\, Mpc$. A smaller value of $\tilde{\alpha} = r_c^2$, as expected from its running behaviour, would give even more pronounced dominance of Einstein--Hilbert term in the inner region.

Inside the Solar system the value of $\alpha$ was calculated some time ago in \cite{Mannheim2007} with a different approach and it was shown that the Solar system tests of gravity are not affected by the existence of the Weyl term. In \cite{Mannheim2007}, it is found that $\alpha = 3.29 \times 10^{75}\ kg\cdot m^2 / sec$, which corresponds to almost 
$r_c \approx 20 kpc$. This value is much smaller than the value obtained for the required phenomenological behaviour in the galaxy. This unconventional running behaviour of $\alpha$ as suggested in \cite{Donoghue2017} that might depend on some physical property of the system besides the energy is evidently seen here. It is imperative that more work is needed to understand the true nature of the coupling $\alpha$.

%%%%%%%%%%%%%%%%%%%%%%%%%%%

\section{Interrelations of Bach tensor components \label{Bmn}} 

The form of the metric solution is given by (\ref{met}) with $ d\Omega_{k}^2 \equiv \frac{1}{1-kx^2}dx^2+(1-kx^2)dy^2 $ corresponding to two dimensional hyperbola, torus and sphere geometries of the solid angle, with $k$ taking values $-1,\, 0,\, 1$, respectively. 

Since Bach tensor has trace and divergence free properties, field equations are not independent. We first note that "xx" and "yy" components of Bach tensor are the same, $B^x_x=B^y_y$. Then the trace of equations (\ref{Bach}) is given by
\begin{equation} \label{b1}
B^r_r + B^t_t +2B^x_x=0.
\end{equation}
Furthermore the divergence of the Bach tensor $\nabla_{\mu}{B^{\mu}_{\nu}}=0$ is calculated to be
\begin{equation} \label{b2}
\left(\frac{d}{dr}+\frac{2+w}{r}\right)B^r_r-\frac{w}{r} B^t_t-\frac{2}{r} B^x_x=0.
\end{equation}
Solving these two equations, we find that $B^x_x$ and $B^t_t$ are given in terms of $B^r_r$ as 
\beqa
B^t_t &=& -\frac{1}{1-w}\left( r\frac{d}{dr}+3+w\right)B^r_r\ , \\
B^x_x &=& \frac{1}{2(1-w)}\left( r\frac{d}{dr}+2(1+w)\right)B^r_r\ .
\eeqa

If a solution satisfies the equation $B^r_r = 0$, then all the field equations (\ref{Bach}) are satisfied by that solution due to above relations.

%%%%%%%%%%%%%%%%%%%%%%%%%%%%%%%%%%%%%%%%%%%%

\end{document}